\documentstyle[aps,prl]{revtex}
\begin{document}
\input epsf.sty
\twocolumn[\hsize\textwidth\columnwidth\hsize\csname %
@twocolumnfalse\endcsname
\draft
\widetext

%
%
%

\title{Uncorrelated and correlated nanoscale lattice distortions in the
paramagnetic phase of magnetoresistive manganites}

\author{V. Kiryukhin$^{1}$, A. Borissov$^1$, J. S. Ahn$^{1}$, Q. Huang$^2$,
J. W. Lynn$^2$,
and S-W. Cheong$^1$}
\address{(1) Department of Physics and Astronomy, Rutgers University,
Piscataway, New Jersey 08854}
\address{(2) NIST Center for Neutron Research, NIST, Gaithersburg, Maryland
20899}

\date{\today}
\maketitle

\begin{abstract}

Neutron scattering measurements on a magnetoresistive manganite
La$_{0.75}$(Ca$_{0.45}$Sr$_{0.55}$)$_{0.25}$MnO$_3$
show that uncorrelated
dynamic polaronic lattice distortions are present in both the
orthorhombic (O) and rhombohedral (R) paramagnetic phases.
The uncorrelated distortions do not exhibit any significant anomaly at the
O-to-R transition. Thus, both the paramagnetic phases are inhomogeneous on the
nanometer scale, as confirmed further
by strong damping of the acoustic phonons and by
the anomalous Debye-Waller factors in these phases.
In contrast, recent x-ray measurements and our neutron data show that
polaronic correlations are present only in the O phase.
In optimally doped manganites,
the R phase is metallic, while the O paramagnetic state is insulating (or
semiconducting). These measurements therefore
strongly suggest that the {\it correlated} lattice
distortions are primarily
responsible for the insulating character of the paramagnetic
state in magnetoresistive manganites.

\end{abstract}


\phantom{.}
]
\narrowtext

\section{Introduction}

Transition metal oxides exhibit a number of unusual phenomena, such as
high-temperature superconductivity, colossal magnetoresistance (CMR), and
phases possessing charge or orbital order. In many of these cases, competition
between various magnetic and electronic ground states gives rise to 
electronically inhomogeneous states, with characteristic length scales 
varying from microns to nanometers \cite{D}. The physical properties of these
states are often sensitive to small changes in the external conditions,
such as an applied magnetic field or pressure, thus giving rise to a number
of the ``colossal'' effects. One of the best known examples of such an
effect is the CMR effect in perovskite manganites $A_{1-x}B_x$MnO$_3$
\cite{Rev}. In these
compounds, application of a magnetic field induces a transition from
a paramagnetic insulating (PI) to a ferromagnetic metallic (FM) state. 
The large difference between the resistivities of these two phases lies at
the heart of the CMR effect. The metallic nature of the FM phase has been
explained within the framework of the double-exchange mechanism \cite{Zener}.
The physical properties of the PI state, however, still remain poorly 
understood.

It is now well established that nanoscale inhomogeneities resulting from
electron-lattice coupling play a key role in the PI state in manganites
\cite{D}. 
There is a significant amount of experimental evidence that small lattice
polarons are present in this state \cite{Sol,SP,Manella,XP}. 
In a simplified model, a lattice polaron forms when 
an $e_g$ electron localizes on a Mn$^{3+}$ ion, and the surrounding
oxygen octahedron distorts due to the Jahn-Teller (JT) effect. Mn$^{4+}$ ions,
which contain three localized $d$ electrons in the $t_{2g}$ orbitals,
are JT inactive, and Mn$^{4+}$O$_6$ octahedra remain undistorted. 
In the vicinity of the ``optimal'' doping $x\approx$0.3, at which the CMR effect
is large, the nominal concentrations of the Mn$^{3+}$ and Mn$^{4+}$ 
ions (1-$x$ and $x$, respectively) are comparable. 
Thus, it is natural to expect that at these dopings
the polarons will
interact with each other through their overlapping long-range
lattice distortions.
Structural correlations attributed to such an interaction have indeed
been observed in neutron and x-ray diffraction experiments in a broad range of
doping levels 0.2$\leq x\leq$0.5 \cite{Corr,Corr1}. These correlations
are very short-range, with a correlation length of several lattice
constants (10-30 $\rm\AA$). It was proposed that the correlated regions exhibit
CE-type and striped charge and orbital ordered structures similar to those
found in manganites possessing long-range charge and orbital order \cite{Corr1}.
However,
some recent experiments are inconsistent with this simple model \cite{Rhombo}, 
and the 
detailed structure of the correlated regions has not been established thus far.

Both correlated and uncorrelated polaronic distortions are believed to 
affect the transport properties of CMR manganites \cite{D}. 
In particular, these
distortions have been associated with the enhanced electrical resistivity of 
the PI state in orthorhombic manganites. It was found that the polaronic
distortions disappear on the transition from the PI to the FM state, which
can be induced either by decreasing temperature, or by an applied magnetic
field. However, the specific roles of the correlated and
uncorrelated distortions in the PI state have not been determined thus far. 

To study these roles, one needs to compare paramagnetic phases
with and without polaronic correlations, preferably in the same sample. 
An elegant way to carry out such a study is made possible by the fact that 
the paramagnetic
state in CMR manganites exhibits two different pseudo-cubic perovskite
structures: an orthorhombic (O) and a rhombohedral (R) 
phase \cite{Rev,Mit,Rad}. 
The MnO$_6$ octahedra in the O phase are distorted, and this distortion is
usually explained by electron-lattice coupling via the Jahn-Teller effect.
The MnO$_6$ octahedra in the average structure of the R phase determined by
powder diffraction, on the other hand, 
are undistorted \cite{Mit}. There is evidence, however,
that dynamic JT distortions 
are still present in the R phase \cite{Manella,Rad,Millis}.
Interestingly, polaronic
correlations have been found thus far only in the 
orthorhombic PI phase \cite{Rhombo}.
Therefore, a sample exhibiting an O-to-R transition in the paramagnetic state 
provides an ideal system for investigation of effects of polaronic correlations
on various physical properties.

In this work, we investigate properties of correlated and uncorrelated polaronic
distortions in La$_{0.75}$(Ca$_{0.45}$Sr$_{0.55}$)$_{0.25}$MnO$_3$
using elastic and inelastic neutron scattering.
This compound undergoes an O-to-R transition within the 
paramagnetic state.
We find that uncorrelated polaronic distortions are present in
both the O and R insulating phases. These distortions are dynamic (quasielastic)
with the characteristic lifetime $\tau_u\approx$120 fs,
and do not show any significant anomaly at the O-to-R transition temperature
$T_s$. Single polarons, therefore, do not appear to play any significant
role in this transition. The polaronic correlations
exhibit a significantly larger lifetime $\tau_c>$600 fs. 
According
to recent x-ray measurements, these correlations are present only in
the O phase, and disappear abruptly at $T_s$ \cite{Rhombo}.
Since structural
changes at $T_s$ are small, we argue that the drop in the electrical 
resistivity at the O-to-R transition
can be explained by the disappearance of the slowly fluctuating
polaronic correlations.
At doping levels near the optimal level of $x\approx$0.3,
the paramagnetic R phase is usually described as metallic, 
while the O state is insulating or semiconducting.
Thus, our data strongly suggest that the {\it correlated} lattice
distortions play the primary role in the microscopic mechanism explaining the
insulating character of the paramagnetic state in the CMR manganites.

\section{Experimental Procedures}

A single crystal of La$_{0.75}$(Ca$_{0.45}$Sr$_{0.55}$)$_{0.25}$MnO$_3$
was grown using the floating zone technique. 
Polycrystalline La$_{0.75}$(Ca$_{0.45}$Sr$_{0.55}$)$_{0.25}$MnO$_3$
was prepared using standard solid state reaction
techniques. 
With increasing temperature, 
La$_{0.75}$(Ca$_{0.45}$Sr$_{0.55}$)$_{0.25}$MnO$_3$
undergoes two transitions: firstly
a FM to orthorhombic PI transition at $T_c\approx$300 K, and then a structural
transition to a paramagnetic rhombohedral state at $T_s\approx$360 K. 
The electrical resistivity shows an abrupt drop of approximately 20\%
at the first-order O-to-R transition \cite{Rhombo}.

Neutron scattering measurements were performed using the BT-2
thermal triple-axis spectrometer at the NIST center for neutron research. 
60$^\prime$-40$^\prime$-S-40$^\prime$-80$^\prime$ 
collimation and fixed final energies of 14.8 meV (inelastic
measurements) and 30.5 meV (elastic measurements) were used. PG filters were
used when appropriate to suppress higher-order wavelength contamination. 
Powder diffraction measurements were performed using the BT-1 powder
diffractometer. A Cu (311) monochromator, $\lambda$=1.5403(2) $\rm\AA$,
and 15$^\prime$-20$^\prime$-S-7$^\prime$ 
collimation were utilized. The samples were mounted
in a closed-cycle refrigerator (30-400 K). 
Crystal structure refinements were carried out using the 
GSAS program \cite{GSAS}. Refinements with isotropic temperature parameters for
all atoms, and with anisotropic parameters for oxygens were performed. 
In this paper, we use
the orthorhombic $Pbnm$ notation, in which the $a$, $b$, and $c$ axes run
along the (1,1,0), (1,-1,0), and (0,0,1) cubic perovskite directions,
respectively. Scattering vectors ($h$,$k$,$l$) are given in reciprocal
lattice units.

\begin{figure}
\centerline{\epsfxsize=2.9in\epsfbox{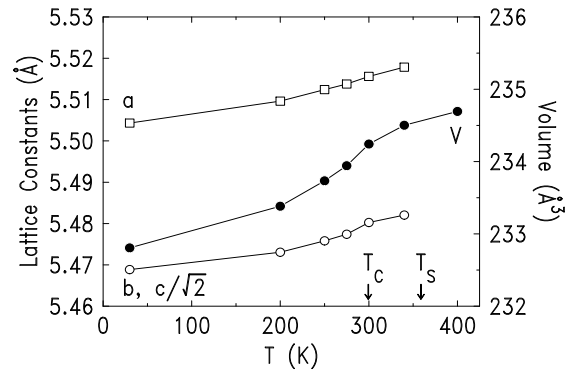}}
\vskip 5mm
\caption{Temperature dependence of the lattice parameters and unit cell volume
in La$_{0.75}$(Ca$_{0.45}$Sr$_{0.55}$)$_{0.25}$MnO$_3$.
The unit cell volume in the R phase ($T$=400 K) is multipled by two.}
\label{fig1}
\end{figure}

\section{Results}
\subsection{Powder Diffraction}

\begin{figure}
\centerline{\epsfxsize=2.9in\epsfbox{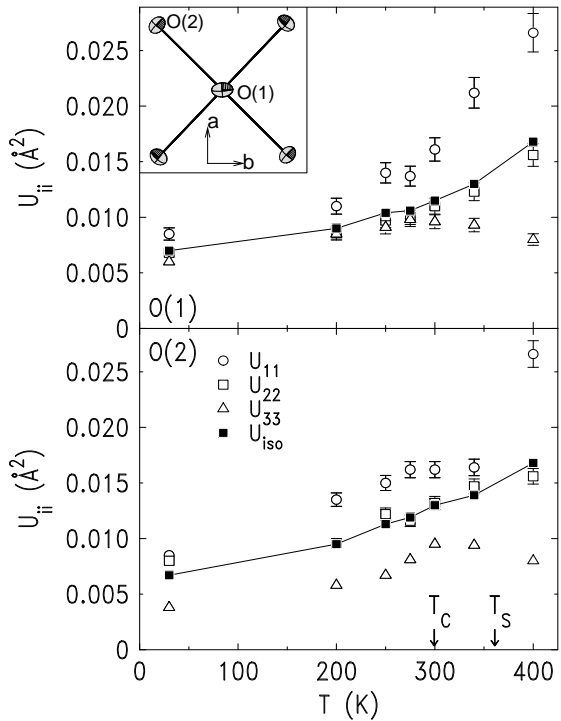}}
\vskip 5mm
\caption{Isotropic ($U_{iso}$) and anisotropic ($U_{ii}$) temperature parameters
for the apical oxygen O(1) and the in-plane oxygens O(2). The anisotropic
parameters $U_{ii}$ represent the major axes of the thermal ellipsoids,
as discussed in the text.
The inset shows projections of the oxygen thermal ellipsoids in a MnO$_6$
octahedron on the $ab$-plane in the O phase at $T$=340 K.}
\label{fig2}
\end{figure}

As shown in the next Section, general crystallographic data are needed to
understand the results of the experiments with single crystals. Therefore,
we first discuss the temperature-dependent parameters of the average structure
obtained in the powder diffraction experiments. Figure \ref{fig1} shows
temperature dependence of the lattice parameters and of the unit cell
volume. The observed anomaly at $T_c$ is consistent with previously reported
results \cite{DW}. 
In Fig. \ref{fig2}, temperature parameters for the apical oxygen
atom O(1) and the in-plane oxygens O(2) are shown (see the inset in Fig.
\ref{fig2} for the explanation of this notation). In a perfect crystal, these
parameters reflect thermal vibrations of the atoms, and therefore grow
with temperature. However, it is well known that in CMR manganites, the
observed temperature dependence of the temperature parameters
cannot be accounted for by the lattice expansion \cite{Rad,DW}. 
According to the 
phenomenological model discussed in Ref. \cite{DW}, the Debye-Waller factors
$W$ in a non-anomalous crystal are related to volume by
$W(T_2)/W(T_1)=[V(T_2)/V(T_1)]^{2\gamma}$. Here 
$W({\bf q})=\langle({\bf q\cdot u})^2\rangle /2$, 
$\langle{\bf u}^2\rangle$ is the rms thermal 
displacement, and $\gamma$ is the Gr\"uneisen constant, which for most materials
is between 2 and 3. Using the values of $U_{iso}$ and $V$ shown in Figs.
\ref{fig1} and \ref{fig2}, we find unphysically large values of the effective
Gr\"uneisen constant $\gamma\approx$50.

Thus, the volume expansion is too small to account for the anomalously 
large change of the thermal parameters $U_{iso}$ with temperature. 
Anomalous enhancement of the thermal parameters in the PI phase is usually
attributed to the dynamic JT effect \cite{Rad,DW}. 
It can also result from presence of
static nanoscale lattice distortions. Irrespective of the exact origin of this
anomalous behavior, it reflects the presence of local imperfections (static or
dynamic) in the crystal structure. We conclude, therefore, that both the O
and R insulating states are inhomogeneous. Importantly, the thermal parameters
continue their anomalous increase above $T_s$. Thus, local lattice distortions,
most likely dynamic JT polarons,
are present in the R state, even though the MnO$_6$ octahedra are undistorted
in the average long-range R structure. 

In addition to the isotropic parameters $U_{iso}$, we show in Fig. \ref{fig2} 
anisotropic temperature parameters $U_{ii}$. To illustrate the true anisotropy
of the thermal ellipsoids, the parameters $U_{ii}$ shown in Fig. \ref{fig2} 
are calculated in the basis of the major axes of these ellipsoids. Thus, 
$U_{ii}$ represent the lengths of the major axes. The exact values of the
$U_{ii}$ often depend on the details of the structural refinement, and 
therefore may be unreliable. However, they may provide valuable 
qualitative information
about the character of the thermal motion in the system. 
Specifically, the data of Fig. \ref{fig2} show that the apical oxygens
O(1) exhibit conventional isotropic thermal displacements at low temperature in
the FM state. However, these displacements quickly become very anisotropic in
the PI state. The inset in Fig. \ref{fig2} shows that the largest thermal
displacements of O(1) at T=340 K occur along the $b$ 
orthorhombic direction. It also
shows that the smallest dispacements of the O(2) in-plane 
atoms are approximately 
collinear with the Mn-O bonds. These data suggest that the JT polarons
are accommodated in the crystal lattice via tilts of the MnO$_6$ octahedra in
which the apical oxygens shift in the direction bisecting the Mn-O-Mn angle.
Interestingly, the large JT distortions of the MnO$_6$ octahedra characteristic
to the static orbital order found in the half-doped manganites are also 
accommodated via staggered tilts in the same direction \cite{Rad2}. 
Combined together,
these observations indicate that the most energetically favorable way to 
accommodate a polaron in the perovskite structure of a CMR manganite is via
the tilt of the MnO$_6$ octahedron in the cubic (110) direction.
Finally, we note that strong anisotropy of the oxygen thermal motions is
also found in the R phase. At this stage, we do not present any simple model 
describing thermal displacements in this phase.

In summary, the above results show that the paramagnetic state exhibits
significant local distortions (static or dynamic) of
the crystal lattice. This is in agreement with previously reported measurements
\cite{Rad,DW}. Interestingly, the local distortions are clearly present in the
R phase, even though the JT distortions are not present in the average 
crystallographic structure. Below, it is shown that the local distortions in 
this phase are associated with the dynamic JT effect. Finally, the atomic 
positions and temperature parameters are determined at various temperatures. 
We use these data to analyze the results discussed in the following Sections. 

\subsection{Single Crystal Experiments}

\begin{figure}
\centerline{\epsfxsize=2.9in\epsfbox{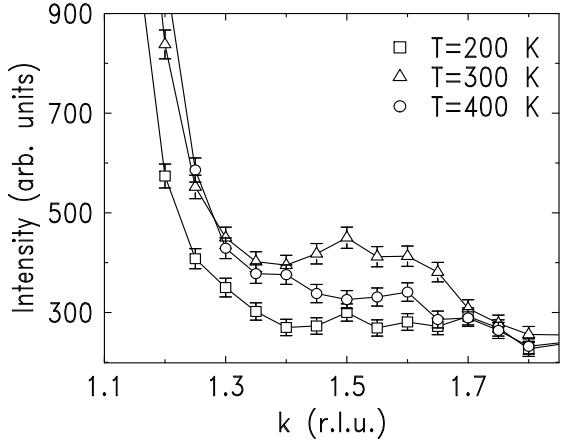}}
\vskip 5mm
\caption{Elastic scans along the (5, $k$, 0) 
direction at various temperatures.}
\label{fig3}
\end{figure}

To investigate the temperature-dependent properties of the correlated
and uncorrelated local lattice distortions, single crystal elastic and inelastic
neutron scattering measurements were carried out.
Figure \ref{fig3} shows elastic
scans in the (5, $k$, 0) direction taken in all the three phases found
in La$_{0.75}$(Ca$_{0.45}$Sr$_{0.55}$)$_{0.25}$MnO$_3$: 
FM ($T$=200 K), orthorhombic paramagnetic ($T$=300 K), 
and rhombohedral paramagnetic
($T$=400 K). A broad peak at $k\approx$1.5 at $T$=300 K is attributed to
polaronic correlations \cite{Corr,Corr1}. 
In agreement with a recent x-ray diffraction study \cite{Rhombo},
we observe this peak only in the orthorhombic phase. For $T>T_c$, additional
scattering intensity centered at $k$=1 (reduced $q$=0) 
and extending over most of the
Brillouin zone is observed. This scattering is conventionally attributed
to scattering from uncorrelated polaronic distortions described as
Jahn-Teller polarons \cite{XP}.

To characterize temperature-dependent dynamic properties of the local
lattice distortions, we have carried out fixed-energy $q$-scans in the
vicinity of $q$=(4, 0.5, 0), and energy scans at the
fixed scattering vector $q$=(4, 0.45, 0) using fixed final energy
$E_f$=14.8 meV. These parameters were chosen to maximize the experimental 
resolution and signal intensity. Unfortunately, elastic $q$-scans were
unobtainable at this configuration because of instrumental limitations.

\begin{figure}
\centerline{\epsfxsize=2.9in\epsfbox{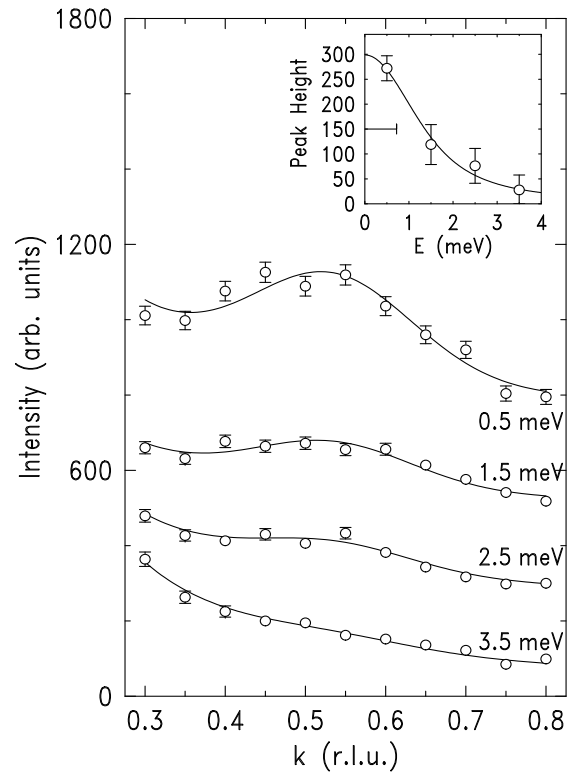}}
\vskip 5mm
\caption{Fixed energy scans along the (4, $k$, 0)         
direction at $T$=300 K. For clarity, the $E$=0.5 meV and 1.5 meV scans
are displaced vertically
by 400 counts, and the $E$=2.5 meV scan -- by 200 counts. The inset shows
the intensity of the peak due to the structural correlations versus energy.
The solid lines are the results of fits, as described in the text. The 
horizontal bar in the inset shows the experimental energy resolution.}
\label{fig4_new}
\end{figure}

\begin{figure}
\centerline{\epsfxsize=2.9in\epsfbox{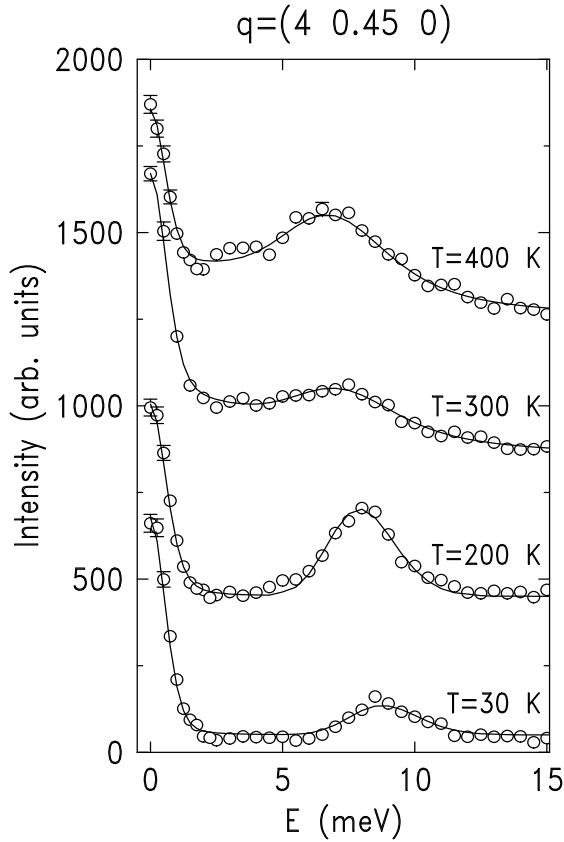}}
\vskip 5mm
\caption{Energy scans at $q$=(4, 0.45, 0) at various temperatures.
The scans are displaced vertically by 400 counts with respect to each other
for clarity. The solid lines are the results of fits, as described in the
text.}
\label{fig4}
\end{figure}

We first address the dynamic properties of the correlated lattice distortions.
Figure \ref{fig4_new} shows scans in the (4, $k$, 0) direction taken at 
various energy transfers $E$. The peak due to the correlated distortions is
clearly observed in the $E$=0.5 meV scan. This scan contains a significant
elastic contribution (the instrumental energy half-width for elastic scattering
is approximately 0.75 meV). The data
of Fig. \ref{fig4_new} were fitted to a sum of a Gaussian line shape and a
monotonically sloping background, the latter described by a power-law 
function. The width and the position of the peak were determined at $E$=0.5 meV
and then kept fixed for higher energies.
The inset in Fig. \ref{fig4_new} shows the intensity of the
correlated peak extracted from these fits at various energy transfers. The data
shown in the inset were fitted to a Lorentzian peak shape convoluted with
the experimental resolution. The solid line in the inset shows the result of
this fit. The obtained intrinsic Lorentzian width for the correlated scattering
is $\Gamma_c$=0.9$\pm$0.15 meV. This fit gives the $E$=0 intensity that is
only 10\% larger than that for $E$=0.5 meV. 
On the other hand, the inelastic scans shown in Fig. \ref{fig4} 
indicate that the intensity for $E$=0 is typically about 30-40\% larger than 
that for $E$=0.5 meV. One can, therefore, speculate that the fit of
Fig. \ref{fig4_new} overestimates the value of $\Gamma_c$.  
Leaving the speculations aside, our measurements set an
upper limit for $\Gamma_c$ at 1.1 meV. Thus, the structural correlations, while
not completely static, exhibit the characteristic lifetime
$\tau_c=\hbar/\Gamma_c$ exceeding 600 fs. 
We note that in bi-layered
manganites, the corresponding lifetime was found to be
an order of magnitude smaller \cite{Lynn}. The correlated regions, therefore, 
fluctuate quite slowly in our sample.

Figure \ref{fig4} shows energy scans at several representative temperatures
at $q$=(4, 0.45, 0). At this $q$, a significant signal from the correlated
distortions is observed (see Fig. \ref{fig4_new}), while no
background due to possible $\lambda$/2 contamination is present.
Also, because of the proximity of a strong Bragg peak (4, 0, 0), a significant
signal due to the uncorrelated distortions should be present
at this $q$. 
For $T<T_c$, the scans of Fig. \ref{fig4}
show only an elastic peak at $E$=0, and a 
resolution-limited transverse acoustic (TA) phonon peak at 
$E\approx$8 meV. This scattering
pattern changes dramatically as temperature is increased above $T_c$. Firstly,
a broad quasielastic scattering signal appears. Secondly, the TA
phonon peak broadens, indicating that the phonons are damped. To analyze 
these data, one therefore should consider contributions from truly elastic
scattering, from the TA phonon, and two quasielastic contributions: one
from the correlated distortions with $\Gamma_c\sim$1 meV, 
and the second from the
broad peak appearing for $T>T_c$. Fits containing all these contributions did
not converge well. Instead, we fitted the data to a sum of a nearly elastic
peak whose Lorentzian width was refined in the fits, a broad Lorentzian
quasielastic peak, and a TA phonon peak. The fits were convoluted with the
instrumental resolution. The obtained fits
are shown as solid lines in Fig. \ref{fig4}. 
In this procedure, the correlated scattering is
effectively included in the nearly-elastic peak, as shown below.
Importantly, the temperature-dependent properties of the broad quasielastic
feature are not affected significantly by the choice of the fitting
algorithm at low energies. This feature is the main subject of 
interest below, and therefore the adopted fitting procedure is
adequate for the purposes of this paper.

\begin{figure}
\centerline{\epsfxsize=2.9in\epsfbox{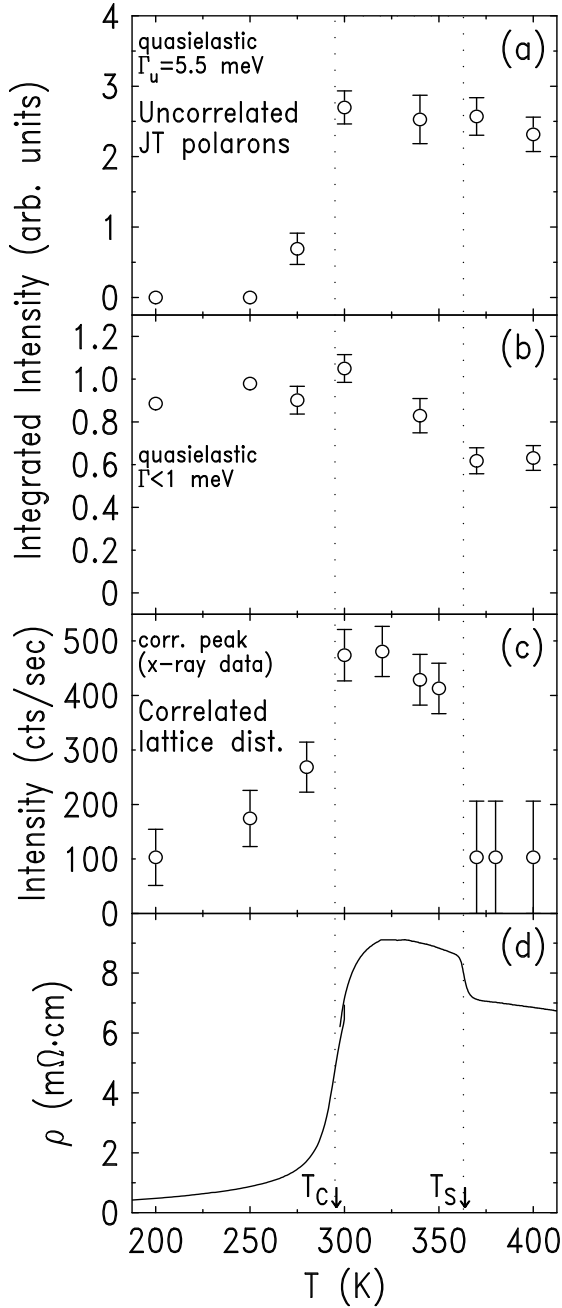}}
\vskip 5mm
\caption{Temperature dependence of the energy-integrated intensity of (a) the
quasielastic scattering due to the uncorrelated polarons ($\Gamma_u$=5.5 meV), 
and (b) the nearly-elastic scattering with $\Gamma<$1 meV 
at $q$=(4, 0.45, 0), extracted
from the data of Fig. 5 as described in the text.
Panel (c) shows temperature dependence of the intensity of the x-ray 
peak due to the 
correlated lattice distortions, and panel (d) -- the electrical resistivity
of the sample. The data shown in (c) and (d) are taken from Ref. [10].}
\label{fig5}
\end{figure}

The Lorentzian width of the broad quasielastic feature determined from the
data of Fig. \ref{fig4} is $\Gamma_u$=5.5$\pm$0.5 meV. 
The energy-integrated intensity of this feature is shown in Fig. \ref{fig5}(a). 
The quasielastic scattering is present only for $T>T_c$. Importantly, it 
does not exhibit any appreciable anomaly at $T_s$. Energy-integrating
x-ray experiments, on the 
other hand, show that the structural correlations abruptly disappear at the 
O-to-R transition \cite{Rhombo}. Thus, the quasielastic intensity shown in
Fig. \ref{fig5}(a) does not contain any significant contribution
from the structural correlations. We therefore conclude that this intensity
originates from the uncorrelated structural distortions. 

The temperature dependence of the nearly-elastic integrated intensity is shown
in Fig. \ref{fig5}(b). The Lorentzian width of this feature was found to be
$\Gamma\lesssim$0.1 meV for $T<T_c$ and $T>T_s$, and $\Gamma\sim$0.2-0.3 meV for
$T_c<T<T_s$. This intensity contains contributions from
the correlated lattice distortions, temperature-dependent incoherent
scattering, as well as a possible contribution due to static uncorrelated 
polarons. This temperature dependence exhibits a certain resemblance to the 
temperature-dependent x-ray intensity of the correlated peak shown in Fig.
\ref{fig5}(c). In particular, the nearly-elastic intensity in the R phase is
smaller than that in the O state. Such a drop in intensity cannot be caused
by the variation in the incoherent scattering. 
Indeed, using the Debye-Waller factors determined in our powder
diffraction measurements, we find \cite{Lov}
that the incoherent scattering changes linearly with temperature,
decreasing by less than 10\% as the temperature increases from 300 K to 400 K.
However, while a certain measurable contribution of the correlated
scattering appears to be present in the data of Fig. \ref{fig5}(b), the
nearly-elastic component is clearly dominated by the background effects.
Nevertheless, the data of Fig. \ref{fig5}(b) can be used to make an estimate
of the upper limit for the integrated intensity due to the correlated 
distortions by measuring the difference between the intensities in the O and
R phases. The combined data of Figs. \ref{fig5} (a) and (b) clearly show that
the integrated intensity due to the dynamic uncorrelated distortions exceeds
this upper limit by at least a factor of 5. The uncorrelated dynamic polarons
are, therefore, the dominant component of the local lattice distortion 
for $T>T_c$.

Finally, we show the intrinsic width $\Gamma_{TA}$ and the energy of the
TA phonon at $q$=(4, 0.45, 0) in Fig. \ref{fig6}. The phonon acquires finite
lifetime for $T>T_c$. As discussed below, this is the result of the presence
of the structural inhomogeneities in the paramagnetic phase.

\section{Discussion}

The above measurements indicate that dynamic
single polarons are the dominant component 
of the local lattice distortion in the paramagnetic state in our sample. 
We cannot 
exclude that some of the uncorrelated polarons are static, but the comparison
of the raw integrated intensities of the elastic and inelastic scattering
(Fig. \ref{fig5}) shows that the large majority of the uncorrelated polarons 
are dynamic.
The characteristic lifetime of the polaronic
distortions is $\tau\sim\hbar/\Gamma_u\approx$120 fs.
Both the O and R paramagnetic phases exhibit
the uncorrelated lattice distortions, and no significant anomaly
in the quasielastic scattering is observed at $T_s$.
The dynamic polarons are absent in the FM phase.

\begin{figure}
\centerline{\epsfxsize=2.9in\epsfbox{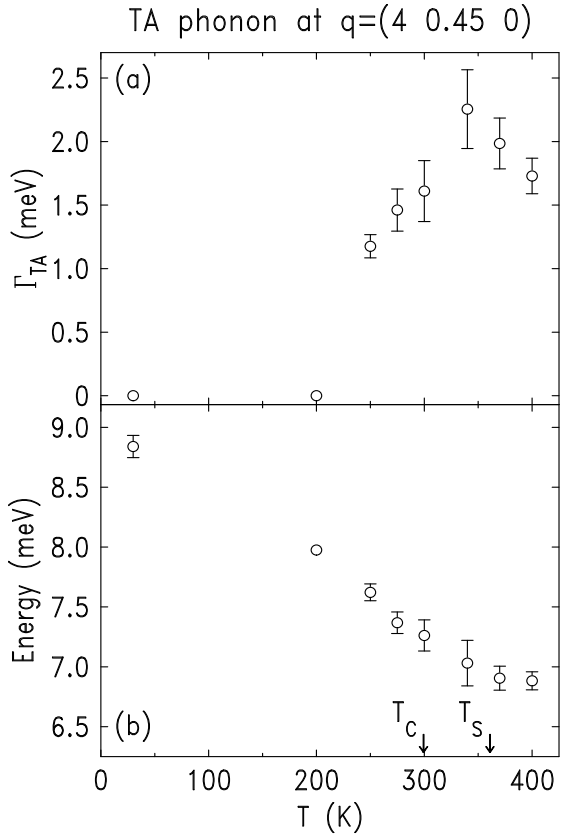}}
\vskip 5mm
\caption{Temperature dependence of the intrinsic Lorentzian width (a), and
the energy (b) of the TA phonon at $q$=(4, 0.45, 0).}
\label{fig6}
\end{figure}

The physical properties showing anomalous behavior in both the O and R
paramagnetic states should, therefore, be associated with the effects of
the {\it dynamic uncorrelated polarons}. 
Firstly, these are the anomalously large
Debye-Waller factors. The second anomalous 
effect observed only in the paramagnetic state is damping of the acoustic
phonons, that acquire a finite lifetime of $\hbar/\Gamma_{TA}\sim$300-400 fs
for $T>T_c$ (see Fig. \ref{fig6}). Because acoustic phonons provide
a major contribution to the thermal conductivity in manganites \cite{T0,Sol},
this damping may be an important factor leading to
its strong reduction in the 
paramagnetic phase \cite{Thermo}.
We note that an even more dramatic damping of the
optical Jahn-Teller
phonon modes in related manganite compounds has been reported
\cite{Lynn,Optical}.
All these effects are, of course,
natural consequences of the presence of local inhomogeneities in
the crystal lattice. In fact, phonon anomalies similar to those found in the
manganites have been seen in several
other systems exhibiting nanoscale inhomogeneities.
In perovskite ferroelectric relaxors \cite{Cross}, for example, anomalous phonon
spectrum and phonon damping effects have been explained by the presence
of nanoscale lattice inhomogeneities described as ``polarized nanoscale
regions'' \cite{RF}. 

Another anomalous property of our sample is a very significant dependence
of the TA phonon energy on temperature in both the ferromagnetic and the
paramagnetic phases (see Fig. \ref{fig6}).
It is unclear whether this effect can be associated with the 
local structural inhomogeneities. Indeed, the temperature dependence
of the phonon energy does not exhibit any significant anomaly at $T_c$, at which
both the correlated and uncorrelated local lattice
distortions appear. 
At this stage, we can only note that the observed phonon softening cannot be
explained by the conventional effects of lattice unharmonicity. Thermal
expansion of a crystalline lattice does give rise to a shift in the
phonon frequency $\Delta(q)=-3\gamma\epsilon\omega(q)$, where $\gamma$ is
the Gr\"uneisen constant, $\epsilon$ is the linear expansivity, and
$\omega(q)$ is the phonon frequency \cite{Lov}. However,
using the data of Figs. \ref{fig1} and
\ref{fig6}, we obtain unphysically large values for the Gr\"uneisen constant
$\gamma\sim$20. The origin of the observed softening of the TA phonon,
therefore, remains to be clarified.

We now discuss the correlated lattice distortions. Compared with the
uncorrelated polarons, these distortions fluctuate very slowly.
Indeed, their characteristic lifetime $\tau_c$ is larger than 600 fs,
thus exceeding the lifetime of the uncorrelated polarons by more than a
factor of 5. As the x-ray data of Fig. \ref{fig5}(c) show, the correlated
distortions are present only in the O phase. 
The increased electrical resistivity $\rho$ of the O phase [Fig. \ref{fig5}(d)]
should therefore be associated
with the presence of the correlated distortions. Indeed, the uncorrelated
polarons do not show any anomaly at $T_s$, and the
differences between the average structures of the O and R phases
are too small to explain the observed changes of $\rho$ at $T_s$.
To provide a quantitative estimate of
the possible effects of the structural change on the electrical resistivity,
we note that the electronic bandwidth in manganites is proportional to
$\cos\omega/d^{3.5}$, where $\omega$=($\pi-\langle$Mn-O-Mn$\rangle$)/2,
$d$ is the Mn-O bond length, and $\langle$Mn-O-Mn$\rangle$
is the Mn-O-Mn angle \cite{Rad3}.
Using the structural data obtained in our powder diffraction measurements,
we find that the difference in the bandwidths of the R and O phases due to
their structural differences is smaller than 0.3\%. While this argument is
not strict in the case of the nonmetallic paramagnetic state, it does show
that the structural changes that take place at $T_s$ have a negligible
effect on the overall electronic band structure. In principle,
it should also be possible to make a numerical estimate for the effects of the
structural correlations on the electrical resistivity \cite{Estimate}. Such
estimates require, however, knowledge of the exact structure of the correlated
regions and their concentration, which are unknown at this stage.

Summing up, it is the presence of the {\it correlated} lattice distortions that
distinguishes the O and R paramagnetic phases in our sample on the 
microscopic level \cite{Rhombo}. 
While both the uncorrelated and the correlated distortions
should contribute to the increased electrical resistivity of the paramagnetic
state, the larger resistivity of the O phase stems from the presence of
the correlated distortions. This is the central result of our work. 

\begin{figure}
\centerline{\epsfxsize=3.8in\epsfbox{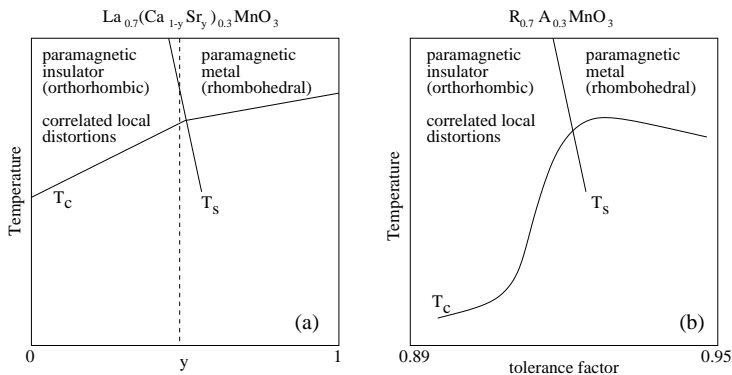}}
\vskip 5mm
\caption{Schematic phase diagrams of optimally doped manganites.
(a) La$_{0.7}$(Ca$_{1-y}$Sr$_y$)$_{0.3}$MnO$_3$ system. The dashed line
indicates the approximate position of our sample in this diagram. (b) General
temperature vs tolerance factor phase diagram. Phases are described as
proposed in the text.}
\label{fig7}
\end{figure}

To understand the significance of this result, we consider
the relationship between the average lattice symmetry and bulk
properties of the ``optimally-doped'' CMR manganites ($x\sim$0.3). Our sample
has composition La$_{1-x}$(Ca$_{1-y}$Sr$_y$)$_x$MnO$_3$ with $x$=0.25 and
$y$=0.55. The phase diagram of this compound family \cite{PD} for $x\sim$0.3 is
sketched in Fig. \ref{fig7}(a).  As $y$ increases from 0 to 1, the average
structure changes from the orthorhombic to rhombohedral in a first-order
transition at $y\sim$0.5. The paramagnetic state is insulating (semiconducting)
in the O phase, while it is metallic on the R side of 
the phase boundary \cite{PD}.
Recent x-ray data \cite{Rhombo} show that the correlated distortions are present
only in the O state. Thus, the insulator-metal transition that the 
{\it paramagnetic} phase undergoes with increasing $y$ is
associated with the destruction of the correlated lattice distortions. 
These data, combined with our characterization of the
uncorrelated distortions and the structural changes at $T_s$, 
strongly suggest that it is the presence of the correlated lattice distortions  
that renders the paramagnetic phase insulating.

Further experiments are needed to establish whether these results can be
generalized to all the perovskite manganites.
We believe, however, that there is sufficient evidence
in hand to warrant a {\it proposal} for the microscopic mechanism 
explaining the observed variation of the properties of the paramagnetic
phase in optimally doped ($x\sim$0.3) manganites.
In Fig. \ref{fig7}(b), a sketch of the general phase diagram of 
$x\sim$0.3 perovskite manganites is shown \cite{Hwang,Mira}. 
As the tolerance factor (defined as $t=d_{Ln-O}/\sqrt 2d_{Mn-O}$) increases,
the system undergoes a structural transition from the orthorhombic to the 
rhombohedral phase. Extensive measurements of bulk properties \cite{Mira}
indicate that electron-lattice coupling effects play a much more significant
role on the orthorhombic side of the phase diagram. A large number of
anomalous properties, such as large volume and magnetic entropy changes
at $T_c$, high magnetostriction, and the largest values of magnetoresistance
(CMR) are observed in the orthorhombic phase. These anomalies are absent or
strongly reduced in the rhombohedral state \cite{Mira}. 
Correlated lattice distortions have been found in a large number of 
orthorhombic perovskite manganites with different compositions 
\cite{Corr,Corr1}. No significant lattice correlations have been reported
thus far in the rhombohedral state \cite{Rhombo}. 
These observations cover a significant part of the phase diagram of Fig.
\ref{fig7}(b), from the narrow-bandwidth
Pr$_{0.7}$Ca$_{0.3}$MnO$_3$ ($t\approx$0.905) to the large-bandwidth
La$_{0.7}$Sr$_{0.3}$MnO$_3$ ($t\approx$0.93). 
In addition, combined results of Ref. \cite{Rhombo} and this work
clearly show that
it is the correlated distortions, and not single polarons, that distinguish
the R and O states on the microscopic level in a sample undergoing the
O-to-R transition in the paramagnetic phase. We believe that these
observations allow us to propose that
the observed drastic differences between the properties of the paramagnetic
O and R states in hole-doped manganites
stem from the presence of the correlated lattice distortions
in the O phase.  

If this proposal is valid, the question arises: does 
the presence of the correlated lattice distortions distinguish
the metallic and the insulating paramagnetic states in manganites? Indeed,
for $x\sim$0.3, the R phase is usually described as a metal, while the O
phase is considered an insulator \cite{Rev}. These conclusions are made on
the basis of the temperature dependence of the electrical resistivity.  
We note that in our sample,  
the resistivity exhibits an insulator-like behavior,
decreasing with increasing temperature in both the O and the R phases,
see Fig. \ref{fig5}(d). One possible explanation for this behavior is that the 
doping level in our sample, $x$=0.25, is lower than the optimal level. 
In addition, the temperature dependence of the resistivity at high temperatures
may not be a good indicator of whether the state can be considered metallic or
not. In fact, it may not even be possible to clearly distinguish a bad metal
from a bad insulator based only on the high-temperature data. A Mott insulator,
for example, shows a continuous crossover from bad metal to bad insulator
with decreasing band width at high temperatures \cite{Kotliar}. Thus,
the scenario
in which the structural O-to-R transition in the paramagnetic state
is associated with an
insulator-metal transition, and in which
the metal is distinguished from the insulator by the absence
of the correlated distortions, clearly deserves further investigation.

Finally, we note that there is a number of basic questions about the role
of nanoscale inhomogeneities in manganites that still remain to be
answered. Firstly, the structure of the correlated regions is still 
unknown \cite{Rhombo}. Secondly, it is unclear why
the correlations are observed only in the O state. A possible explanation
will likely invoke the effects of the favorable elastic response of the O
structure to the long-range strains \cite{lrs}
produced by the local distortions. 
Even though interesting theoretical results on the role of long-range
strains in manganites have recently been achieved \cite{D2,Ahn},
realistic models taking into account the long-range elastic effects
are yet to be developed. 
Thus, while clear connection between the 
nanoscale structural
inhomogeneities and the bulk properties in manganites has been now established,
further work is needed to determine the nature of these inhomogeneities and
the physical mechanism of their formation.

\section{Conclusions}

We present a neutron scattering investigation of the paramagnetic
phase in La$_{0.75}$(Ca$_{0.45}$Sr$_{0.55}$)$_{0.25}$MnO$_3$. 
The obtained results can be summarized as
follows. (i) Both the orthorhombic (O) and the rhombohedral (R) paramagnetic
phases exhibit dynamic single polarons. The polarons strongly affect
the dynamics of the crystal lattice, as evidenced by the anomalous 
Debye-Waller factors and damped acoustic phonons in the paramagnetic phase.
The damping of the acoustic modes may be responsible for the reduced thermal 
conductivity characteristic of this phase. (ii) Nanoscale structural 
correlations fluctuate much more slowly than the uncorrelated polarons. 
In agreement with previous x-ray diffraction measurements \cite{Rhombo}, 
the correlations are observed only in the orthorhombic phase. Thus,
it is the presence of the structural correlations that distinguishes the O
phase from the R phase on the microscopic level. (iii) Single polarons do 
not show any sharp anomaly at the O-to-R transition, and the structural
changes at this transition are not large enough to affect the electrical
properties significantly. Thus, the increased electrical resistivity of
the O phase should be attributed to the presence of the correlated
lattice distortions. (iv) Based on these results, as well as on previously
published work, we propose that the insulating
character of the orthorhombic paramagnetic phase in CMR manganites 
stems from the presence of the {\it correlated} lattice distortions.

\section{Acknowledgements}
We are grateful to J. P. Hill for important discussions.
This work was supported by the NSF under Grant No. DMR-0093143, and by the
NSF MRSEC program, Grant No. DMR-0080008.


\end{document}